\documentclass[12pt]{article}
\usepackage{amsmath}

\newcommand{\bm}{\begin{multiline}}
\newcommand{\beq}{\begin{equation}}
\newcommand{\eeq}{\end{equation}}
\newcommand{\beqs}{\begin{eqnarray}}
\newcommand{\eeqs}{\end{eqnarray}}
\newcommand{\tr}{\mbox{\rm Tr}}
\newcommand{\ra}{\rightarrow}

\begin{document}

\thispagestyle{empty}

\begin{flushright}
gr-qc/yymmxxx\\
\end{flushright}

\hfill{}

\hfill{}

\hfill{}

\vspace{32pt}

\begin{center}

\textbf{\Large Accelerating Taub-NUT and Eguchi-Hanson solitons in four dimensions}

\vspace{48pt}

{\bf Brenda Chng,}\footnote{E-mail: \tt{phycmyb@nus.edu.sg}}
{\bf Robert Mann}\footnote{E-mail: 
{\tt mann@avatar.uwaterloo.ca}}
{\bf and Cristian Stelea}\footnote{E-mail: 
{\tt cistelea@uwaterloo.ca}}

\vspace*{0.2cm}

{\it $^{1}$Department of Physics, National University of Singapore}\\
{\it 2 Science Drive 3, Singapore 117542}\\[.5em]

{\it $^{2}$Perimeter Institute for Theoretical Physics}\\
{\it 31 Caroline St. N. Waterloo, Ontario N2L 2Y5 , Canada}\\[.5em]

{\it $^{2,3}$Department of Physics, University of Waterloo}\\
{\it 200 University Avenue West, Waterloo, Ontario N2L 3G1, Canada}\\[.5em]

\end{center}

\vspace{30pt}

\begin{abstract}
 We construct new solutions of the vacuum Einstein field equations in four dimensions via a solution generating method utilizing the SL$(2,R)$ symmetry of the reduced Lagrangian. We apply the method to an accelerating version of the Zipoy-Voorhees solution and generate new solutions which we interpret to be the accelerating versions of the Zipoy-Voorhees generalisation of the Taub-NUT solution (with Lorentzian signature) and the Zipoy-Voorhees generalisation of the Eguchi-Hanson solitons (with Euclidean signature). As an intermediary in the solution-generating process we obtain charged versions of the accelerated Zipoy-Voorhees-like families of solutions. Finally we present the accelerating version of the Taub-NUT solution and discuss its properties.
\end{abstract}

\vspace{32pt}

\setcounter{footnote}{0}

\newpage

\section{Introduction}

Ever since the formulation of General Relativity, exact solutions have played an integral part in our understanding of the nature of spacetime. For example, much of our understanding of black hole thermodynamics and inflation were possible only with the discovery of the Kerr-Newman and 
FRW solutions respectively. Given the importance of such exact solutions, there is a corresponding impetus to derive new solutions which upon analysis would yield further insight into our universe. Since Einstein's equations in their unadulterated form consist of a series of coupled non-linear differential equations, obtaining solutions by hand is intractable unless some kind of simplifying symmetry is imposed in the ansatz. This motivated the development of many ingenious and powerful strategies utilizing such simplifications to derive solutions to Einstein's equations. Of particular importance for the present work is a special type of simplifying ansatz, the \textit{static} axisymmetric Weyl-Papapetrou metric, which was first proposed by Weyl in \cite{weyl1}  %
\begin{eqnarray}\label{weyl}
ds_4^2&=&-e^{-\psi}dt^2 +e^{\psi}\big[e^{2\mu}(d\rho^2+dz^2)+\rho^2d\varphi^2\big].
\end{eqnarray}
The metric is specified by the values of two functions $\psi$ and $\mu$, which are functions of the canonical Weyl variables $\rho$ and $z$. Starting with a static vacuum axisymmetric solution of the Einstein field equations, consider its dimensional reduction along the timelike direction down to three dimensions. The reduced Lagrangian can be written as:
\beqs
{\cal L}_3=eR-\frac{1}{2}e(\partial\psi)^2,
\eeqs
where we denote $e=\sqrt{g}$ and $R$ is computed with the three-dimensional metric:
\beqs
ds^2&=&e^{2\mu}(d\rho^2+dz^2)+\rho^2d\varphi^2.
\label{3dim}
\eeqs
The equation of motion for $\psi$ is then readily seen to be $\Delta\psi=0$, where $\Delta$ is the Laplacian constructed using the three-dimensional metric (\ref{3dim}). Now, the key observation is that $\Delta\psi=e^{-2\mu}\Delta_{\mu=0}\psi$, where $\Delta_{\mu=0}$ is the Laplacian computed for a flat three-dimensional Euclidean metric, which corresponds to setting $\mu=0$ in (\ref{3dim}). Therefore any  solution $\psi(\rho,z)$ of Laplace's equation in the flat three dimensional space is automatically a valid solution of Laplace's equation in the curved background (\ref{3dim}). Once we know $\psi$,  the remaining function $\mu(\rho,z)$ is found by performing a simple line-integral using the relations: 
\begin{eqnarray}\label{gammap1}
\partial_z{\mu}=\frac{\rho}{2}\partial_{\rho} \psi\partial_z \psi,~~~~~~~\partial_{\rho}{\mu}=
\frac{\rho}{4}\left[\left(\partial_{\rho} \psi\right)^2-\left(\partial_{z} \psi\right)^2\right].
\end{eqnarray}
The Einstein's equations in $4$-dimensions for a static axisymmetric background are now essentially reduced to finding a solution of Laplace's equation on flat space. Due to the linearity of the Laplace equation for $\psi$, construction of multi-black hole versions is easily carried out. The Weyl formalism has been recently extended to higher dimensions by Emparan and Reall \cite{emparan&reall} and the same line of thought can be used for the corresponding higher dimensional axisymmetric metrics.

Given the simplifications introduced by the above axisymmetric ansatz, one now has two choices. One may either try to solve the differential equations directly, or, in more general cases, try to further exploit the hidden symmetries of the dimensionally reduced Lagrangians and generate solutions using pre-existing solutions as seeds. 

Consider for instance the following `scaling' symmetry of the field equations: given a vacuum static solution described by the pair of functions $(\psi, \mu)$ then it is easily seen from (\ref{gammap1}) that the pair $(\gamma\psi, \gamma^2\mu)$ will describe new vacuum static axisymmetric solution of the field equations, where $\gamma$ is any real parameter. As an example, its application converts the Schwarzschild solution into the Zipoy-Voorhees solution \cite{Zipoy}. This simple observation will be liberally utilised in the present work.

There also exist transformations similar to the Ehlers-Harrison transformation for the Ernst formalism \cite{ehlers-harrison,ernst1,Stephani:tm} which map static vacuum solutions into stationary Einstein-Maxwell solutions. For the Weyl-Papapetrou ansatz, it has been long known that a transformation already exists that brings a static, axisymmetric vacuum solution to a non-trivial class of static solutions in Einstein-Maxwell theory \cite{weyl2}. In particular, the Schwarzschild solution can be transformed into  the Reissner-Nordstr\"{o}m  solution. In this paper, we will demonstrate a simpler alternative derivation of this transformation using a $SL(2,R)$ symmetry of the reduced Lagrangian in three dimensions.\footnote{A similar idea has been considered in \cite{Yazadjiev}, however the details of the charging transformation differ here.} However, unlike previous applications of this transformation, we show that combining this transformation with the above scaling symmetry, we are able to generate new solutions. Our seed metric will be an accelerating version of the Zipoy-Voorhees solution \cite{Teo:2005wf}. In particular, we obtain new vacuum stationary axisymmetric metrics which we interpret as describing the accelerating Zipoy-Vorhees-like family of Taub-NUT solutions (with Lorentzian signature) and Eguchi-Hanson instantons (with Euclidean signature). Much like the original Zipoy-Voorhees solution \cite{Zipoy}, such metrics are parameterized by a real number $\alpha$. For $\alpha=1$ we recover the accelerating Taub-NUT/Eguchi-Hanson solitons and, for higher positive integer values of $\alpha$, they can be interpreted as the `superposition' of accelerating $\alpha$ NUT-charged objects/solitons. 

The structure of this paper is as follows. In section \ref{method} we describe our solution generating technique, which maps a static axisymmetric solution in vacuum to a new stationary vacuum solution of  Einstein's gravity in four dimensions. In section \ref{cmetricseedsection} we apply the transformation technique on accelerating solutions, namely on the C-metric and on the accelerating Zipoy-Voorhees metric and we consider more closely the properties of the generated accelerating Taub-NUT solution in section \ref{properties}. Finally, in section \ref{conclusions} we summarise the main results of this paper and discuss potential avenues for further research. In Appendix A we cast the accelerating Taub-NUT metric in Weyl-Papapetrou form.

\section{Weyl's charging method: the $SL(2,R)$ approach}
\label{method}

We start with Einstein-Maxwell theory in four dimensions described by the Lagrangian:
\beqs\label{4dEMlagrangian}
\mathcal{L}_{4}=eR-\frac{1}{4}eF^{2}_{(2)}, 
\eeqs
where $R$ is the Ricci scalar computed with the $4$-dimensional metric, $F_{(2)}=dA_{(1)}$ is the electromagnetic field strength which only has an electric component $A_{(1)}=\chi dt$ and we denote $e=\sqrt{-g}$. Consider the dimensional reduction of the four-dimensional Lagrangian (\ref{4dEMlagrangian}) to three dimensions on a timelike coordinate using the static Kaluza-Klein ansatz:
\begin{eqnarray}
ds_{4}^2 =  -e^{-\phi}dt^2+e^{\phi}ds_{3}^2.
\label{4dto3d}
\end{eqnarray}
The reduced Lagrangian in three dimensions is then\footnote{Here $e=\sqrt{|g|}$ and $R$ is computed with the $3$-dimensional metric $ds^2_3$.}
\begin{eqnarray}
\mathcal{L}_{3} &=&eR-\frac{1}{2}e(\partial {\phi })^{2}+\frac{1}{2}ee^{\phi}(\partial {\chi })^{2}.
\label{3dlag1}
\end{eqnarray}
Notice now that if one defines the matrix:
\beqs 
\cal{M}=\left(\begin{array}{cc}  e^{\frac{\phi}{2}} & \frac{\chi}{2} e^{\frac{\phi}{2}} \\
 \frac{\chi}{2} e^{\frac{\phi}{2}} & -e^{-\frac{\phi}{2}}+\frac{\chi^2}{4}e^{\frac{\phi}{2}} \end{array}\right),
\eeqs
then the three-dimensional Lagrangian can be cast into the following form:
\beqs
{\cal L}_{3}&=&eR+e\tr\big[\partial{\cal M}^{-1}\partial{\cal M}\big],
\eeqs
The reduced Lagrangian is then manifestly invariant under general $SL(2,R)$ transformations if one considers the following transformation laws for the three-dimensional fields:
\beqs
g_{\mu\nu}\ra g_{\mu\nu},~~~~{\cal M}\ra\Omega^T{\cal M}\Omega,~~~~~~~\Omega=\left(\begin{array}{cc}a&b\\ c&d\end{array}\right), ~~~~ad-bc=1.
\eeqs
Starting now with a static axisymmetric vacuum solution described by the metric:
\beqs
ds^2&=& -e^{-\psi}dt^2+e^{\psi}ds_{3}^2,
\eeqs
then performing the dimensional reduction on the timelike direction down to three dimensions and applying a general $SL(2,R)$ transformation parameterized as above, we obtain a static axisymmetric electrically charged solution of Einstein-Maxwell field equations, described by the fields:
\beqs
ds^2&=&-e^{-\phi}dt^2+e^{\phi}ds_3^2, ~~~~~~~A_{(1)}=\chi dt,\nonumber\\
e^{\phi}&=&e^{\psi}\frac{\left(1-\delta e^{-\psi}\right)^2}{4C^2\delta},~~~~~~~~
\chi=\frac{4C\delta}{e^{\psi}-\delta},
\label{Echarging}
\eeqs
where, in terms of the parameters appearing in $\Omega$, the new constants $\delta$ and $C$ can be expressed as $\delta=c^2/a^2$ and $C=1/(2ac)$. Note that in the limit in which $\Omega=I_2$, \textit{i.e.} $c\ra0$ and $a=1$, we have $\delta\ra0$ simultaneously with $C\ra\pm\infty$ such that the product $C^2\delta\ra 1/4$ remains constant.

As an example of this charging technique, let us generate the Reissner-Nordstr\"om solution starting from the Schwarzschild metric:
\beqs
ds^2&=&-\left(1-\frac{2m}{r}\right)dt^2+\frac{dr^2}{1-\frac{2m}{r}}+r^2(d\theta^2+\sin^2\theta d\varphi^2).
\label{schw}
\eeqs
The final solution can be written in the form:
\beqs
ds^2&=&-\frac{4C^2\delta r(r-2m)}{\left((1-\delta)r+2\delta m\right)^2}dt^2+\frac{\left((1-\delta)r+2\delta m\right)^2}{4C^2\delta r(r-2m)}dr^2+\frac{\left((1-\delta)r+2\delta m\right)^2}{4C^2\delta}(d\theta^2+\sin^2\theta d\varphi^2),\nonumber\\
A_{(1)}&=&\frac{2C\left((1+\delta)r-2\delta m\right)}{(1-\delta)r+2\delta m}dt.
\eeqs
For generic values $\delta\neq 1$ we can easily perform a redefinition of the radial coordinate, together with an appropriate constant scaling of the timelike coordinate and cast the solution into the usual Reissner-Nordstr\"om form. The electric charge is $Q=m/C$, while the mass of the solution is $M=(1+\delta)Q/(2\sqrt{\delta})$.  Note that $\delta=1$ is a special case as it leads to the Bertotti-Robinson metric \cite{Bertotti} and it therefore describes the extremally charged Reissner-Nordstr\"om solution for which $M=Q$.

At this point let us note that this method is not restricted only to metrics with axisymmetric symmetry; it can be extended to any general static vacuum solution of Einstein's field equations. Also, one can easily consider a similar method to generate magnetically charged static solutions out of axial vacuum metrics. 

Once we obtained a static electrically charged solution in Einstein-Maxwell theory, the next step in our solution-generating technique will be to perform a dualisation of the electromagnetic potential and find the corresponding magnetically charged solutions. In our case it turns out that it is easier to compute the dual electromagnetic potential in the reduced three-dimensional theory. In this case we start with the three-dimensional Lagrangian  (\ref{3dlag1}) and dualise the scalar field to obtain a magnetic $2$-form field strength $F_{(2)}$. Following the usual dualisation procedure, we add a term $d\chi\wedge F_{(2)}$ to the action and solve the equations of motion for the scalar field $\chi$. Replacing the result in the action we finally express the Lagrangian in terms of the dual field as:\footnote{Note that we perform the dualisation using a three dimensional \textit{Euclidean} metric.}
\beqs
{\cal L}_{3}&=&eR-\frac{1}{2}e(\partial\phi)^2-\frac{1}{4}ee^{-\phi}F_{(2)}^2,
\eeqs
where the components of the two-form field strength are computed using the formula:
\beqs
F_{\alpha\beta}&=&ee^{\phi}\epsilon_{\alpha\beta\mu}\partial^{\mu}\chi,
\label{Fdual}
\eeqs
where $\epsilon_{\alpha\beta\mu}$ is the Levi-Civita symbol. After lifting the solution back to four-dimensions we obtain a magnetically charged static solution of the Einstein-Maxwell field equations. 

We are now ready to perform the last step in our solution-generating method, namely to map the magnetic solution to a vacuum axisymmetric stationary solution of Einstein's field equations in four dimensions. This actually involves two steps: we first map the magnetic solution of the Einstein-Maxwell theory to a solution of the Einstein-Maxwell-Dilaton (EMD) theory with a specific value of the dilaton coupling, namely the one corresponding to the Kaluza-Klein theory, \textit{i.e.} $a=-\sqrt{3}$. To do this we shall employ the general results derived in \cite{Emparan:2001bb} (see also \cite{Brenda} for a geometrical derivation of the respective mapping). Starting with a magnetostatic solution:
\beqs
ds_4^2&=&-e^{-\phi}dt^2+e^{\phi}\big[e^{2\mu}(d\rho^2+dz^2)+\rho^2 d\varphi^2\big], ~~~~~~~A_{(1)}=A_{\varphi} d\varphi,
\eeqs 
the corresponding solution of the EMD system is:
\beqs
ds_4^2&=&-e^{-\frac{\phi}{4}}dt^2+e^{\frac{\phi}{4}}\bigg[\left(e^{2\mu}\right)^{\frac{1}{4}}(d\rho^2+dz^2)+\rho^2 d\varphi^2\bigg],\nonumber\\A_{(1)}&=&\frac{A_{\varphi}}{2} d\varphi, ~~~~e^{\frac{\tilde{\phi}}{\sqrt{3}}}=e^{\frac{\phi}{4}}.
\eeqs
This is none other than the dimensional reduction of a vacuum five-dimensional metric using the ansatz:
\beqs
ds_5^2&=&e^{-\frac{2\tilde{\phi}}{\sqrt{3}}}(dz+\frac{A_{\varphi}}{2}d\varphi)^2+e^{\frac{\tilde{\phi}}{\sqrt{3}}}ds_4^2.
\eeqs
In our case it turns out that the $5$-dimensional metric is simply the trivial product of a $4$-dimensional Euclidean metric with a time direction. Since the $5$-dimensional metric solves the vacuum Einstein equations it is manifest that the $4$-dimensional Euclidean metric will be Ricci flat, \textit{i.e.} it solves the vacuum Einstein equations in four dimensions. Therefore, our final result is expressed in the form:
\beqs
ds_4^2&=&e^{-\frac{\phi}{2}}(dz+A_{\varphi}/2 d\varphi)^2+e^{\frac{\phi}{2}}\bigg[\left(e^{2\mu}\right)^{\frac{1}{4}}(d\rho^2+dz^2)+\rho^2 d\varphi^2\bigg].
\label{finalEuclid}
\eeqs

We note that even if the charging method does not require any other symmetry beyond the static condition, this second step in our solution-generating technique can be applied \textit{only} to stationary axisymmetric metrics that can be cast into the Weyl-Papapetrou form. However, this is not really a very stringent constraint as most of the physically interesting solutions can be cast in the Weyl-Papapetrou form. To understand the effects of this last step in our solution-generating method one could take for instance the magnetically charged four-dimensional Reissner-Nordstr\"om and map it to an Euclidean vacuum metric as in (\ref{finalEuclid}). However, given the presence of the square roots appearing in the factors $e^{\frac{\phi}{2}}$ it is easily seen that we obtain an axisymmetric NUT-charged solution with naked singularities, whose physical interpretation is obscure at this point. On the other hand, since we restrict ourselves to axisymmetric metrics that can be cast in Weyl form, it turns out that before we apply the charging procedure one can use the scaling symmetry discussed in introduction to scale the dilaton in the initial seed in such a way to cancel the awkward effect of the square-roots in the final expression of the metric.
  
\section{Accelerating Zipoy-Voorhees-like families of solutions}
\label{cmetricseedsection}
In four dimensions, a particularly interesting class of solutions that generalise the Schwarzschild black-hole is the so-called C-metric. The static part of this metric was found by Levi-Civita almost one century ago (see \cite{Stephani:tm}), however, its physical interpretation was clarified only after Kinnersley and Walker's work decades later \cite{Kinnersley:1970zw}. By performing appropriate coordinate definitions, they found that this metric describes a pair of causally disconnected black holes uniformly accelerating in opposite directions. The cause of the acceleration is understood in terms of nodal/conical singularities along the axis that connects the two black holes and these singularities are interpreted as strings/struts pulling or pushing the black holes apart. A more general class of electrovacuum spacetimes that includes and considerably generalises the C-metric was found by Pleba\'nski and Demia\'nski \cite{Plebanski:1976gy} (see also in \cite{ray} the general Type D family of solutions in EM theory). Recent analyses of this class of solutions have been performed in \cite{hongteo,griffiths&podolsky}, where a new exact solution describing a pair of accelerating and rotating charged black holes having also a NUT-charge has been presented. However, an accelerating NUT solution \textit{without rotation} has not been identified yet within that class \cite{griffiths&podolsky}. It is the goal of this section to try to construct such an accelerating solution. We shall consider next the uncharged C-metric, respectively the accelerating Zipoy-Voorhees solution \cite{Teo:2005wf} as the seeds in our solution-generating procedure.

Expressed in the form given in \cite{hongteo} the C-metric takes the simple form:
\beqs
ds^2&=&\frac{1}{A^2(x-y)^2}\bigg[-(y^2-1)F(y)dt^2+\frac{dy^2}{(y^2-1)F(y)}+\frac{dx^2}{(1-x^2)F(x)}+(1-x^2)F(x)d\varphi^2\bigg],\nonumber
\eeqs
where $F(\xi)=1+2mA\xi$. We restrict our attention to case in which $0\leq 2mA<1$ and, in order to preserve the signature of the metric we restrict the values of the coordinates such that:
\beqs
-\frac{1}{2mA}\leq y \leq -1,~~~~~~~~-1\leq x\leq 1.
\eeqs
In terms of these coordinates, spatial infinity corresponds to $x=y=-1$, the black hole horizon is located at $y=-\frac{1}{2mA}$, while acceleration horizon corresponds to $y=-1$. The part of the symmetry axis joining the black hole horizon with the acceleration horizon is $x=1$, while the one joining the black hole horizon to infinity is $x=-1$. 

In order to apply our solution generating technique we need to write the C-metric in Weyl form. Using the results from \cite{hongteo} we obtain:
\beqs
ds^2&=&-e^{-\psi}dt^2+e^{\psi}\big[e^{2\mu}(d\rho^2+dz^2)+\rho^2d\varphi^2\big],\nonumber\\
e^{-\psi}&=&\frac{(y^2-1)F(y)}{A^2(x-y)^2},~~~~~e^{2\mu}=\frac{(y^2-1)F(y)}{f(x,y)G(x,y)},
\eeqs
where 
\beqs
f(x,y)&=&(y^2-1)F(x)+(1-x^2)F(y),\nonumber\\
G(x,y)&=&\big[1+mA(x+y)^2\big]^2-m^2A^2(1-xy)^2,
\eeqs
while the canonical Weyl coordinates $\rho$ and $z$ are defined such that:
\beqs
\rho^2&=&\frac{(y^2-1)(1-x^2)F(x)F(y)}{A^4(x-y)^4},~~~~~z=\frac{(1-xy)[1+mA(x+y)]}{A^2(x-y)^2},\nonumber\\
d\rho^2+dz^2&=&\frac{f(x,y)G(x,y)}{A^4(x-y)^4}\left(\frac{dy^2}{(y^2-1)F(y)}+\frac{dx^2}{(1-x^2)F(x)}\right).
\label{weylc}
\eeqs
Consider now the scaling transformation $(\psi, \mu)\ra(\gamma\psi, \gamma^2\mu)$, where $\gamma$ is a real parameter. Applying it to the C-metric we obtain a new vacuum solution of the form:
\beqs
ds^2&=&-\bigg[\frac{(y^2-1)F(y)}{A^2(x-y)^2}\bigg]^{\gamma}dt^2+\big[A^2(x-y)^2]^{\gamma-2}\left[\frac{(1-x^2)F(x)}{[(y^2-1)F(y)]^{\gamma-1}}d\varphi^2\right.\nonumber\\
&&+\frac{\left((y^2-1)F(y)\right)^{\gamma^2-\gamma}}{\big[f(x,y)G(x,y)\big]^{\gamma^2-1}}\left(\frac{dy^2}{(y^2-1)F(y)}+\frac{dx^2}{(1-x^2)F(x)}\right)\bigg].
\label{ZipoyC}
\eeqs
It is clear that by taking $\gamma=1$ we
recover the initial C-metric. On the other hand, let us consider
the zero-acceleration limit of this metric. Performing the
coordinate transformations: \beqs
x&=&\cos\theta,~~~y=-\frac{1}{Ar},~~~t\ra A^{2\gamma-1}t, \eeqs
while taking the limit $A\ra 0$ \textit{and} rescaling the metric
by a constant factor $A^{2\gamma-2}$ it is readily seen that we
recover the Zipoy-Voorhees solution (\cite{Zipoy}). Therefore, one
could naively interpret the metric (\ref{ZipoyC}) as describing an
accelerating version of the Zipoy-Voorhees solution. However,  the
fact that the above metric is not the `proper' accelerating Zipoy-Voorhees
solution can also be seen from the fact that the
$\gamma=2$ of this family should reduce to the so-called
accelerating Darmois solution.
This is the coincident limit of the accelerating Bonnor dihole solution that was recently found by Teo
in \cite{Teo:2005wf}. In fact, a different metric describing the accelerated Zipoy-Voorhees solution, however, written in a very symmetric form has been presented by Teo in the same work.
The `proper' accelerating Zipoy-Voorhees solution reads: \beqs
ds^2&=&-e^{-\psi}dt^2+e^{\psi}\big[e^{2\mu}(d\rho^2+dz^2)+\rho^2d\varphi^2\big],\\
e^{-\psi}&=&\frac{(y^2-1)F(y)}{A^2(x-y)^2}\left(\frac{F(y)}{F(x)}\right)^{\alpha-1},~~e^{2\mu}=\frac{(y^2-1)F(y)}{f(x,y)G(x,y)}\frac{F(y)^{\alpha^2-1}F(x)^{(\alpha-1)^2}}{G(x,y)^{\alpha^2-1}},\nonumber
\eeqs
where the canonical Weyl coordinates are again defined in (\ref{weylc}). Indeed, we see that the Darmois solution (\textit{i.e.} the $\alpha=2$ member of this family) is clearly different from the $\gamma=2$ member of (\ref{ZipoyC}) and therefore we cannot actually interpret (\ref{ZipoyC}) as being an accelerating version of the Zipoy-Voorhees family.\footnote{We thank Edward Teo for pointing this out to us.} Nonetheless, since the metric (\ref{ZipoyC}) is only an intermediate result in our solution-generating technique, we shall not further discuss its properties at this point, but limit ourselves to notice that one can apply the scaling transformation on Teo's solution and further generate a new family of vacuum metrics indexed by two distinct real parameters:
\beqs
ds^2&=&-e^{-\psi}dt^2+e^{\psi}\big[e^{2\mu}(d\rho^2+dz^2)+\rho^2d\varphi^2\big],\label{ZEd}\\
e^{-\psi}&=&\bigg[\frac{(y^2-1)F(y)}{A^2(x-y)^2}\left(\frac{F(y)}{F(x)}\right)^{\alpha-1}\bigg]^{\gamma},~e^{2\mu}=\left(\frac{(y^2-1)F(y)}{f(x,y)G(x,y)}\frac{F(y)^{\alpha^2-1}F(x)^{(\alpha-1)^2}}{G(x,y)^{\alpha^2-1}}\right)^{\gamma^2},\nonumber
\eeqs

The next step is to charge the solution (\ref{ZEd}) using a general $SL(2,R)$ transformation. Using the formulae
(\ref{Echarging}) we obtain: \beqs
ds^2&=&-e^{-\psi}\frac{1}{H_{\gamma}(x,y)}dt^2+e^{\psi}H_{\gamma}(x,y)\big[e^{2\mu}(d\rho^2+dz^2)+\rho^2d\varphi^2\big],\nonumber\\
A_{(1)}&=&\frac{4C\delta}{\left(\frac{A^2(x-y)^2}{(y^2-1)F(y)}\left(\frac{F(x)}{F(y)}\right)^{\alpha-1}\right)^{\gamma}-\delta}dt,~~~~~H_{\gamma}(x,y)=\frac{\left(1-\delta\left(\frac{(y^2-1)F(y)}{A^2(x-y)^2}\left(\frac{F(y)}{F(x)}\right)^{\alpha-1}\right)^{\gamma}\right)^2}{4C^2\delta}.\nonumber
\eeqs
Let us consider a few limiting cases of the above metric. Taking $\delta\ra0$ and $C\ra\infty$ while keeping the product $C^2\delta$ constant we recover the uncharged metric (\ref{ZEd}). On the other hand, the $\gamma=1$ member of this family should correspond to the charged accelerating Zipoy-Voorhees solution (see also \cite{Richterek:2004bb}). In particular, for $\alpha=1$ this should reduce to a charged version of the C-metric. However, unlike the known form of the electrically charged C-metric, in general, the above solution has a curvature singularity located at the roots of $H_{\gamma}(x,y)=0$. Therefore its interpretation as a new form of the charged C-metric is dubious. 

By dimensionally reducing this solution down to three dimensions and dualising the scalar field $\chi$ to an electromagnetic field as described in section \ref{method}, we find that the magnetic potential is given by:
\beqs
A_{\varphi}&=&\frac{\gamma}{C}\frac{(1-x^2)\left(\alpha F(x)+(1-\alpha)F(y)\right)}{A^2(x-y)^2}+\frac{2m\gamma \alpha x}{AC},
\eeqs
while the metric remains unchanged in this process. Finally, taking $\gamma=2$ and using (\ref{finalEuclid}) we find:
\beqs
ds^4&=&\frac{(y^2-1)F(x)}{A^2(x-y)^2}\left(\frac{F(y)}{F(x)}\right)^{\alpha}\frac{C^2\delta}{H(x,y)}(dt+A_{\varphi}d\varphi)^2+\frac{H(x,y)}{A^2(x-y)^2}\left[\left(\frac{F(x)}{F(y)}\right)^{\alpha}(1-x^2)F(y)d\varphi^2\right.\nonumber\\
&&+\frac{\left(F(x)F(y)\right)^{\alpha(\alpha-1)}}{G(x,y)^{\alpha^2-1}}\left(\frac{dy^2}{(y^2-1)F(y)}+\frac{dx^2}{(1-x^2)F(x)}\right)\bigg],
\label{ZipoyACEuclid} \eeqs where we defined: \beqs
H(x,y)&=&\frac{1-\delta\left(\frac{(y^2-1)F(x)}{A^2(x-y)^2}\left(\frac{F(y)}{F(x)}\right)^{\alpha}\right)^{2}}{2}.
\eeqs This is the main result of this section. In the limit
$\delta\ra0$, $C\ra\infty$ (with $C^2\delta$ constant and
rescaling the $t$ coordinate by a constant factor) we recover the
Euclidean form of the accelerating Zipoy-Voorhees solution
(\ref{ZEd}).  It is manifest that the C-metric corresponds to
the $\alpha=1$ member of this family. Another interesting limit to
consider is the zero-acceleration limit. In this case it turns out
that performing the coordinate transformations: 
\beqs
x&=&\cos\theta,~~~~~y=-\frac{1}{Ar},~~~~~\delta\ra\delta
A^4,~~~~~t\ra\frac{t}{A}, 
\eeqs 
we obtain a general family of vacuum Euclidean solutions, indexed by a real parameter $\alpha$,\footnote{Notice that $C$ is not an essential parameter and it can be absorbed by a constant rescaling of the $z$ coordinate.} which interpret as the Zipoy-Voorhees generalisation of the Eguchi-Hanson soliton: 
\beqs
ds^2&=&\frac{1-\delta \left(1-\frac{2m}{r}\right)^{2\alpha}}{\left(1-\frac{2m}{r}\right)^{\alpha}}\bigg[\left(\frac{r(r-2m)}{(r-m)^2-m^2\cos^2\theta}\right)^{\alpha^2-1}(dr^2+r(r-2m)d\theta^2)+r(r-2m)\sin^2\theta d\varphi^2\bigg]\nonumber\\&&+\frac{4C^2\delta \left(1-\frac{2m}{r}\right)^{\alpha}}{1-\delta \left(1-\frac{2m}{r}\right)^{2\alpha}}(dz+\frac{2m\alpha}{C}\cos\theta d\varphi)^2.
\label{genEH}
\eeqs
Indeed, if we take $\alpha=1$ and properly rescale the $z$ coordinate to absorb some constant factor we obtain a spherically-symmetric metric:
\beqs
ds^2&=&\delta\frac{r(r-2m)}{r^2-\delta(r-2m)^2}(dz+4m\cos\theta d\varphi)^2+\left(r^2-\delta(r-2m)^2\right)\bigg[\frac{dr^2}{r(r-2m)}+d\Omega^2\bigg].\nonumber
\eeqs
We distinguish now two possibilities. If  we set directly $\delta=1$ we can cast the metric in the following form:
\beqs
ds^2&=&\frac{R^2}{4}\left(1-\frac{\sigma^4}{R^4}\right)(dz+\cos\theta d\varphi)^2+\frac{dR^2}{1-\frac{\sigma^4}{R^4}}+\frac{R^2}{4}d\Omega^2,\nonumber
\eeqs
after redefining $R^2=4m(r-m)$ and $\sigma=4m$. This is the well-known Eguchi-Hanson soliton \cite{Eguchi:1978xp}. On the other hand, if $\delta\neq 1$ then by redefining the radial coordinate such that $R^2-n^2=r^2-\delta(r-2m)^2$ with $(1-\delta)n^2=4m^2\delta$ and rescaling $z$ we obtain:
\beqs
ds^2&=&\frac{\left(R+\frac{n}{\sqrt{\delta}}\right)\left(R+n\sqrt{\delta}\right)}{R^2-n^2}(dz+2n\cos\theta d\varphi)^2+\frac{R^2-n^2}{\left(R+\frac{n}{\sqrt{\delta}}\right)\left(R+n\sqrt{\delta}\right)}dR^2+(R^2-n^2)d\Omega^2,\nonumber
\eeqs
which we recognise as the Euclidean version of the Taub-NUT metric. It is now possible to set again $\delta=1$ and recover the extremal Taub-NUT solution. 

Another case of interest is the one that corresponds to negative values for $\delta$. Setting $\delta\ra-\delta$, from the general form of the metric (\ref{genEH}), we obtain a metric with Lorentzian signature:\footnote{We also changed the notation $z\ra t$.}
\beqs
ds^2&=&\frac{1+\delta \left(1-\frac{2m}{r}\right)^{2\alpha}}{\left(1-\frac{2m}{r}\right)^{\alpha}}\bigg[\left(\frac{r(r-2m)}{(r-m)^2-m^2\cos^2\theta}\right)^{\alpha^2-1}(dr^2+r(r-2m)d\theta^2)+r(r-2m)\sin^2\theta d\varphi^2\bigg]\nonumber\\&&-\frac{4C^2\delta \left(1-\frac{2m}{r}\right)^{\alpha}}{1+\delta \left(1-\frac{2m}{r}\right)^{2\alpha}}(dt+\frac{2m\alpha}{C}\cos\theta d\varphi)^2.
\label{genTN}
\eeqs  
where now $\delta$ takes positive values only. Consider now the $\alpha=1$ member of this family of solutions. After redefining the radial coordinate such that $R^2+N^2=r^2+\delta(r-2m)^2$, where $(1+\delta)N^2=4\delta m^2$, we obtain:
\beqs
ds^2&=&-\frac{\left(R\pm\frac{N}{\sqrt{\delta}}\right)\left(R\mp N\sqrt{\delta}\right)}{R^2+N^2}(dt+2N\cos\theta d\varphi)^2+\frac{R^2+N^2}{\left(R\pm\frac{N}{\sqrt{\delta}}\right)\left(R\mp N\sqrt{\delta}\right)}dR^2+(R^2+N^2)d\Omega^2,\nonumber
\eeqs
\textit{i.e.} the Taub-NUT solution \cite{Taub} with mass $M=\pm\frac{N(\delta-1)}{2\sqrt{\delta}}$ and NUT charge $N$. On the other hand, setting $\delta\ra0$ and taking the limit $C\ra\infty$ while keeping the product $C^2\delta$ constant, the metric (\ref{genTN}) is readily seen to reduce to the Zipoy-Voorhees  metric.

 Therefore, we expect that (\ref{ZipoyACEuclid}) describes the accelerating version of the family (\ref{genEH}). Computing some of the curvature invariants for this metric one  finds that generically there is a curvature singularity located at the roots of $H(x,y)=0$ as long as $x\neq y$.  However if we consider negative values of $\delta$ (\textit{i.e.} replace $\delta\ra-\delta$) in the above metric we obtain a vacuum solution with Lorentzian signature and furthermore, we find that $H(x,y)>0$ always (for $x\neq y$).

\section{Properties of the accelerating Taub-NUT solution}
\label{properties}

In what follows we will concentrate our attention on the $\alpha=1$ member of this Lorentzian family. The metric becomes:
\beqs
ds^4&=&-\frac{(y^2-1)F(y)}{A^2(x-y)^2}\frac{C^2\delta}{H(x,y)}\left(dt+\frac{1}{C}\left(\frac{(1-x^2)F(x)}{A^2(x-y)^2}+\frac{2m x}{A}\right)d\varphi\right)^2\nonumber\\
&&+\frac{H(x,y)}{A^2(x-y)^2}\left[(1-x^2)F(x)d\varphi^2+\frac{dy^2}{(y^2-1)F(y)}+\frac{dx^2}{(1-x^2)F(x)}\right],
\label{ZipoyACLorentz}
\eeqs
where we denote:
\beqs
H(x,y)&=&\frac{1+\delta\left(\frac{(y^2-1)F(y)}{A^2(x-y)^2}\right)^{2}}{2}.\nonumber
\eeqs

Let us first notice that the C-metric, respectively the Taub-NUT metric are included as limiting cases in the above solution. Indeed, taking $\delta\ra0$ and $C\ra\infty$ while keeping the product $C^2\delta$ constant, after rescaling the time coordinate with a constant factor we obtain the uncharged C-metric solution. On the other hand, the zero-acceleration limit is taken by performing the coordinate redefinitions and scalings of the parameters:
\beqs
x&=&\cos\theta,~~~y=-\frac{1}{Ar},~~~\delta\ra A^4\delta,~~~t\ra\frac{t}{A},
\eeqs
in the limit $A\ra0$. It is readily seen that in this limit we obtain the Taub-NUT metric.

To understand the properties of this solution it turns out to be more convenient to consider the above metric in Weyl form:
\beqs
ds^2&=&-e^{-\phi}(dt+A_{\varphi}d\varphi)^2+e^{\phi}\big[e^{2\mu}(d\rho^2+dz^2)+\rho^2d\varphi^2\big],\nonumber\\
e^{-\phi}&=&e^{-\psi}\frac{2C^2\delta}{1+\delta e^{-2\psi}},~~~~~e^{2\mu}=\frac{(y^2-1)F(y)}{f(x,y)G(x,y)},
\label{weylfinal}
\eeqs
where the canonical Weyl coordinates $\rho$ and $z$ are given by (\ref{weylc}) and the expressions of $e^{-\psi}$ and $e^{2\mu}$ in terms of the canonical Weyl coordinates are given in Appendix A. The general analysis of the above metric can now be done in parallel with the one corresponding to the uncharged C-metric. In particular, we see that if we restrict the values of $m$ and $A$ such that $0\leq2mA<1$, then, in order to preserve the correct signature of the metric, the coordinates $(x, y)$ have to take the range:
\beqs
-1\leq x\leq 1,~~~~-\frac{1}{2mA}\leq y\leq -1.
\eeqs
Now, it is well known that, in Weyl cylindrical coordinates, black hole horizons correspond
to ÒrodsÓ on the symmetry axis \cite{Emparan:2001bb}. Our interpretation of the above solution as describing an accelerating NUT-charged black hole will rely on the identification of such rods on the symmetry axis.

We will define the symmetry axis to correspond to $\rho=0$,
\textit{i.e.} it is the $z$-axis. Using (\ref{weylc}) one  sees
that it corresponds to the four intervals: $x=-1$, $x=1$,
$y=-\frac{1}{2mA}$ and $y=-1$. As we shall prove bellow, $y=-\frac{1}{2mA}$ corresponds to the event horizon of the black hole, $y=-1$ is the acceleration horizon, the line
$x=1$ is the part of the symmetry axis between the event horizon
and the acceleration horizon, while $x=-1$ is the part of the
symmetry axis joining up the event horizon with asymptotic
infinity.

To this end, notice that the asymptotic region $x=y=-1$ corresponds to $z=\pm\infty$, while the end-points of the range of the coordinates $(x, y)$ are mapped into $z(x, y)$ as follows:
\beqs
z_1&=&z\left(-1, -\frac{1}{2mA}\right)=-\frac{m}{A},~~~~~z_2=z\left(1,-\frac{1}{2mA}\right)=\frac{m}{A},~~~~~z_3=z(1, -1)=\frac{1}{2A^2}.\nonumber
\eeqs

Note now that $e^{-\phi}|_{\rho=0}$ vanishes at all the above points $z_i$, $i=1.. 3$, it is positive for $z<z_1$ and $z_2<z<z_3$ whereas both $e^{-\phi}|_{\rho=0}$ and $e^{2\mu}|_{\rho=0}$ are \textit{zero} for $z_1<z<z_2$ and $z>z_3$. We may then follow a similar analysis with the one performed in \cite{KillingHor} to conclude that the regions $z_1<z<z_2$ and $z>z_3$ are the Killing horizons of our accelerating solution. One can also see this by noting that the location of the horizons is given by the equation $g^{yy}=0$, which in our case corresponds to the equation $(y^2-1)F(y)=0$. Furthermore, by computing the area of each of the above horizons one can check that $z_1<z<z_2$ has finite area and it corresponds then to a black hole horizon, while $z>z_3$ has infinite area and it corresponds to an accelerating horizon. Indeed, using the C-metric coordinates the area of the black hole horizon is readily found to be:
\beqs
A_H&=&\int^{2\pi}_0\int_{-1}^1\sqrt{g_{\varphi\varphi}g_{xx}}dxd\varphi=\frac{8\pi m^2}{C\sqrt{\delta}(1-4m^2A^2)}.
\eeqs
while the area corresponding to $y=-1$ diverges.

Having determined that the above solution describes an accelerating object, let us turn now to a consideration of the `cause' of the acceleration. The analysis of the conical singularities proceeds exactly as in the case of the uncharged C-metric. In particular, if we denote the periodicity of $\varphi$ as $\Delta\varphi$, then along a portion of the axis where the metric function $e^{-\phi}$ is positive, $e^{-\phi}>0$, the deficit of conical angle will be given by\footnote{The measurement of the proper circumference and proper radius must be performed in a frame for which the proper time $d\tau=dt+A_{\varphi}d\varphi=0$.}:
\beqs
\Delta&=&2\pi-\Delta\varphi e^{-\mu}|_{\rho=0}.
\eeqs
Recall that if $\Delta<0$ one has an excess of conical angle and this corresponds to a strut, if $\Delta>0$ one has a deficit of conical angle that corresponds to a string, whereas if $\Delta=0$ there is no conical singularity on that part of the symmetry axis. Since in our case the function $e^{2\mu}$ is precisely the same with the one corresponding to the uncharged C-metric, we deduce that in general there is a conical singularity residing in this solution and that, for appropriate values of $\Delta\varphi$, it can be chosen to lie along $z_2<z<z_3$ (\textit{i.e.} $x=1$) \textit{or} $z<z_1$ (\textit{i.e.} $x=-1$). In particular we find:
\beqs
\Delta_{x=\pm1}&=&2\pi-(1\pm2mA)\Delta\varphi.
\eeqs
One can remove the conical singularity on the segment $z_2<z<z_3$ (\textit{i.e.} $x=1$) if one chooses $\Delta\varphi=2\pi/(1+2mA)$ but then there will be a positive deficit angle for $z<z_1$ (\textit{i.e.} $x=-1$) and this can be interpreted as an semi-infinite cosmic string pulling on the black hole. Alternatively, for $\Delta\varphi=2\pi/(1-2mA)$ one can eliminate the conical angle for $z<z_1$ (\textit{i.e.} $x=-1$) but then there will be a excess of conical angle for $z_2<z<z_3$ (\textit{i.e.} $x=1$). This is interpreted as a strut pushing on the black hole. The strut continues past the acceleration horizon and connects with the mirror black hole on the other side of it.\footnote{The existence of the second black hole on the other side of the acceleration horizon is obscured by the use of the Weyl coordinates.}

Following the discussion in \cite{hongteo} let us consider next
the presence of torsion singularities. In general these appear
when the conical singularities possess a non-zero angular
velocity, signified by a non-vanishing
$\omega=g_{t\varphi}/g_{tt}$ along the symmetry axis. As  is
apparent from metric written in Weyl-Papapetrou form, near the
symmetry axis $\rho\ra0$ for a non-zero value of $\omega$ the
coordinate $\varphi$ will become a timelike coordinate and this
will lead to the apparition of CTCs sufficiently close to the
axis. In general, these CTCs can be eliminated only when $\omega$
takes the \textit{same} constant value along the entire axis of
symmetry as in that case it is possible to perform a global
coordinate transformation $t\ra t-\omega|_{\rho=0}\varphi$ to give
a metric without such pathologies. For our accelerating NUT solution we find that on the symmetry axis $\rho=0$:
\beqs
 \omega|_{x=\pm1}&=&\pm\frac{2m}{AC},
 \eeqs
and therefore at the first sight there are  unavoidable torsion
singularities associated with this metric. However, one can still
perform a coordinate definition $t_N=t+\frac{2m}{AC}\varphi$ on
the line $x=1$ respectively $t_S=t-\frac{2m}{AC}\varphi$ near
$x=-1$. Since the coordinate $\varphi$ is periodic, this will
introduce a periodicity for the time coordinate. However, this is precisely what is expected in the
case of a NUT-charged solution \cite{Misner}.

\section{Conclusions}
\label{conclusions}
In this paper we constructed new solutions of the vacuum Einstein field equations in four dimensions via a solution generating method utilizing the SL(2, R) symmetry of the dimensionally reduced 
action in three dimensions. Our method was based on the simple observation that a static axisymmetric metric as written in Weyl-Papapetrou exhibits a simple `scaling' symmetry that allows one to generate a family of new static vacuum axisymmetric solutions, indexed by a real parameter. In particular, using this scaling symmetry one can easily generate the Zipoy-Voorhees solution from the Schwarzschild solution. 

We also made use of a charging method for static vacuum metrics, which dates back to Weyl \cite{weyl1}. We demonstrated a simpler alternative derivation of this transformation by using a $SL(2,R)$ symmetry of the reduced Lagrangian in three dimensions. However, unlike previous applications of this transformation, we showed that with our simplified mapping and by combining this charging method with the scaling property, one is able to generate new vacuum stationary axisymmetric metrics. The Lorentzian version of the generated solutions gives a Zipoy-Voorhees like generalisation of the accelerating Taub-NUT solutions, while the Euclidean version gives a non-trivial two-parameter generalisation of the Eguchi-Hanson solitons in four dimensions. We focused our attention on a particular member of this family and we showed that it describes the accelerated version of the Taub-NUT space.

As avenues for further research, it would be interesting to study in more detail the connection of the singular charged C-metric that we obtained with the usual form of the C-metric. In particular, it would be interesting to find a proper dilatonic generalisation of the accelerated Zipoy-Voorhees metric, one that would reduce to the proper charged C-metric in the appropriate limit.

\vspace{10pt}

{\Large Acknowledgements}

C.S. would like to thank Edward Teo for valuable remarks on the manuscript. B.C. would like to thank Kenneth Hong for his comments on the draft.

The work of R.B.M. and C.S. was supported by the Natural Sciences and Engineering Research Council of Canada. 

\renewcommand{\theequation}{A-\arabic{equation}} 
\setcounter{equation}{0} 

\section*{A: The Weyl form of the accelerating Taub-NUT metric}
\label{Weyl form}

Following Emparan and Reall \cite{emparan&reall} we introduce the notation:
\beqs
\zeta_i&=&z-z_i,~~~~~R_i=\sqrt{\rho^2+\zeta_i^2},~~~~~Y_{ij}=\rho^2+R_iR_j+\zeta_i\zeta_j.
\eeqs
It can then be shown \cite{emparan&reall} that:
\beqs
R_1-\zeta_1&=&\frac{(y^2-1)F(x)}{A^2(x-y)^2},~~~~~~~
R_1+\zeta_1=\frac{(1-x^2)F(y)}{A^2(x-y)^2},~~~~~~~
R_2-\zeta_2=\frac{(x-1)(y+1)F(x)}{A^2(x-y)^2},\nonumber\\ \nonumber\\
R_2+\zeta_2&=&-\frac{(x+1)(y-1)F(y)}{A^2(x-y)^2},~~~~~~~R_3-\zeta_3=\frac{(x-1)(y+1)F(y)}{A^2(x-y)^2},\nonumber\\
R_3+\zeta_3&=&-\frac{(x+1)(y-1)F(x)}{A^2(x-y)^2},
\eeqs
while:
\beqs
Y_{12}&=&\frac{(x-1)(y-1)(2mA-1)^2}{2A^4(x-y)^2},~~~~~
Y_{13}=\frac{(x-1)(y-1)F(x)F(y)}{2A^4(x-y)^2},~~~~~
Y_{23}=\frac{2F(x)F(y)}{A^4(x-y)^2}.\nonumber
\eeqs

Then the Weyl form of the uncharged C-metric corresponds to the following expressions:
\beqs
e^{-\psi}&=&\frac{(R_1-\zeta_1)(R_3-\zeta_3)}{R_2-\zeta_2},\nonumber\\
e^{2\mu}&=&\frac{1}{4(2mA-1)^2R_1R_2R_3}\frac{Y_{12}Y_{23}}{Y_{13}}\frac{(R_1-\zeta_1)(R_3-\zeta_3)}{R_2-\zeta_2},
\eeqs
from which we can readily find $e^{-\phi}$ in (\ref{weylfinal}). Finally, expressing $x$ in terms of $\rho$ and $z$ we find \cite{hongteo}:
\beqs
A_{\varphi}&=&\frac{1}{C}\left(\frac{(R_1+\zeta_1)(R_2-\zeta_2)}{R_3-\zeta_3}+\frac{2m}{A}\frac{F_1+F_2}{2F_0}\right),
\eeqs
where:
\beqs
F_0&=&4m^2AR_1+m(1+2mA)R_2+m(1-2mA)R_3,\nonumber\\
F_1&=&-4mR_1-2m(1+2mA)R_2+2m(1-2mA)R_3,\nonumber\\
F_2&=&\frac{2m}{A^2}(1-2m^2A^2).
\eeqs
This completes the derivation of the Weyl-Papapetrou form of the accelerated Taub-NUT solution.

\end{document}